\documentclass[aps,prd,preprint,tightenlines,superscriptaddress,
amsfonts,amssymb,amsmath,byrevtex,showpacs,nofootinbib]{revtex4-1}
\usepackage[polish,english]{babel}
\usepackage[applemac]{inputenc}
\usepackage[T1]{fontenc}
\usepackage{latexsym}
\usepackage{graphicx}
\usepackage{geometry}
\usepackage{epstopdf}
\usepackage{subfig}
\usepackage{color}

\usepackage{xcolor}
\usepackage{pdfsync}
\usepackage{fancyhdr}
\usepackage{makeidx}
\usepackage[breaklinks,colorlinks,backref]{hyperref}
\hypersetup{
    colorlinks, 
    linktoc=all, 
    linkcolor=red,  
  citecolor=hyptxt,
  urlcolor=blue}
  \hypersetup{
  citebordercolor=,
  filebordercolor=green,
  linkbordercolor=blue
}
\definecolor{hyptxt}{rgb}{0.7, 0.4, 0.9}
\usepackage{bibtopic}
\newcommand{\bea}{\begin{eqnarray}}
\newcommand{\eea}{\end{eqnarray}}

\newcommand{\dR}{\mathbb R}

\newcommand{\be}{\begin{equation}}
\newcommand{\ee}{\end{equation}}

\newcommand{\I}{\mathbb I}

\newcommand{\R}{\mathbb R}

\newcommand{\ket}[1]{|\kern.3ex#1\kern.3ex\rangle}
\newcommand{\bra}[1]{\langle\kern.3ex #1 \kern.3ex|}
\newcommand{\scalar}[2]{\langle\kern.3ex #1 \kern.3ex|\kern.3ex#2\kern.3ex\rangle}
\newcommand{\norm}[1]{\|\kern.3ex#1\kern.3ex \|}

\newcommand{\Group}[1]{\mathrm{#1}} 
\newcommand{\Bra}[1]{\langle #1 \vert} 
\newcommand{\Ket}[1]{\vert #1 \rangle} 
\newcommand{\BraKet}[2]{\langle #1 \vert #2 \rangle} 

\begin{document}

\title{Robustness of the quantum BKL scenario}

\author{Andrzej G\'{o}\'{z}d\'{z}}
\email{andrzej.gozdz@umcs.lublin.pl}
\affiliation{Institute of Physics, Maria Curie-Sk{\l}odowska
University, pl.  Marii Curie-Sk{\l}odowskiej 1, 20-031 Lublin, Poland}

\author{W{\l}odzimierz Piechocki} \email{wlodzimierz.piechocki@ncbj.gov.pl}
\affiliation{Department of Fundamental Research, National Centre for Nuclear
  Research, Pasteura 7, 02-093 Warszawa, Poland}


\date{\today}

\begin{abstract}
  Quantum Belinski-Khalatnikov-Lifshitz scenario presents an unitary evolution
  of the system.  However, the affine coherent states quantization applied to
  the quantization of the underlying classical scenario depends on the choice of
  the group parametrization. Using two simplest parameterizations of
  the affine group, we show that qualitative features of our quantum system do
  not depend on the choice. It means that the quantum bounce replacing singular
  classical scenario is expected to be a generic feature of considered system.
  This paper complements our recent article \cite{Gozdz:2018aai}.

\end{abstract}


\maketitle


\section{Introduction}
Recently, we have found that the affine coherent states, ACS, quantization
depends on the parametrization of the affine group \cite{AWT}. Since our paper
\cite{Gozdz:2018aai} concerning the quantization of the
Belinski-Khalatnikov-Lifshitz (BKL) scenario is based on the ACS quantization,
it is reasonable to examine the dependence of the results on the group
parametrization. This is the main motivation of the present paper. To have
analytical results, as in previous paper \cite{Gozdz:2018aai}, we consider the
second, the most popular parametrization of the affine group.

It is worth to recall that the BKL scenario concerns the generic singularity of
general relativity (see \cite{BKL22,BKL33,Bel} and \cite{ryan,Nick}). The
resolution of the singularity at the quantum level is of the primary importance
for the quantum gravity programme. Recently, we have found \cite{Gozdz:2018aai}
that the quantum BKL scenario presents an unitary process so that singular
classical BKL evolution is replaced by regular quantum bounce.  However, the
quantization method we have applied is not unique (as any quantization
scheme). Thus, the examination of the robustness of obtained results is an
important issue that cannot be omitted.

\section{Classical dynamics}

For self-consistency of the present paper, we first recall the main results of
Ref. \!\cite{Gozdz:2018aai}.

The two form $\Omega$ defining the Hamiltonian formulation, devoid of the
dynamical constraints, is given by
\begin{equation}\label{rob1}
\Omega = dq_1 \wedge dp_1 + dq_2 \wedge dp_2 + dt \wedge dH \, ,
\end{equation}
where the variables $\{q_1, q_2,p_1, p_2\}$ parameterise the phase space, $H$ is
the Hamiltonian generating the dynamics, and where $t$ is an evolution parameter
(time) corresponding to the specific choice of $H$.  The Hamiltonian reads
\begin{align}
 \nonumber H(t, q_1, q_2, p_1, p_2)
& := - q_2 - \ln \left[-e^{2 q_1}-e^{q_2 - q_1}-\frac{1}{4}(p_1^2
+ p_2^2 +t^2) +\frac{1}{2}(p_1 p_2 + p_1 t + p_2 t)\right]   \\
&\label{rob2} =:  - q_2 - \ln F(t, q_1,q_2,p_1,p_2 )\, ,
\end{align}
where $F(t, q_1,q_2,p_1,p_2 ) > 0.$

The examination of the topology of the phase space and well definiteness of
the logarithmic function in \eqref{rob2} requires
\cite{Gozdz:2018aai}: (i) $(p_1,p_2) \in \dR_+^2$, where $\dR_+ := \{p \in
\R~|~p>0\}$, and (ii) $p_1 \rightarrow 0$ and $p_2 \rightarrow 0$ implies $t
\rightarrow 0^+$.  Thus, considered gravitational system evolves away from the
singularity at $t = 0$.  The range of the variables $q_1$ and $q_2$ results from
the physical interpretation ascribed to them \cite{Czuchry:2012ad} so that
$(q_1, q_2)\in \dR^2$. Thus, the physical phase space $\Pi$ consists of the two
half planes:
\begin{equation}\label{kps2}
\Pi = \Pi_1 \times \Pi_2 :=
\{(q_1, p_1) \in \dR \times \dR_+\} \times
\{(q_2, p_2) \in \dR \times \dR_+\} \; .
\end{equation}
It is important to notice that only the subspace
\begin{equation} \label{AvailablePhaseSpace}
\tilde{\Pi} =\{(q_1,p_1,q_2,p_2)
\colon F(t,q_1,p_1,q_2,p_2)>0 \} \subset \Pi
\end{equation}
is available to the dynamics. It is due to the logarithmic function in the
expression defining the Hamiltonian. To make this restriction explicit, we
rewrite the Hamiltonian \eqref{rob2} in the form
\begin{eqnarray} \label{vis1a}
&& H(t, q_1, q_2, p_1, p_2) =
\begin{cases}
- q_2 - \ln F(t, q_1, q_2, p_1, p_2),~~\text{for}~~F(t, q_1, q_2, p_1, p_2) >0 \\
0,~~~\text{for}~~~F(t, q_1, q_2, p_1, p_2) < 0 \,
\end{cases}
\end{eqnarray}
with $\lim_{F\rightarrow 0^-} H = 0~~\text{and}~~\lim_{F\rightarrow 0^+} H = +
\infty $.

\section{Hilbert space and quantum observables}

Each $\Pi_k~(k=1,2)$ can be identified with the manifold of the affine group
$G:= \Group{Aff}(\dR)$ acting on $\dR$, which is sometimes denoted as
``$px+q$''. In the case considered in \cite{Gozdz:2018aai} the actions of this
group on $\dR_+$ is defined to be
\begin{equation}\label{robb2}
x'=(\tilde{q},\tilde{p})\cdot x =
\tilde{p} x + \tilde{q},~~\mbox{ where }~~
(\tilde{q},\tilde{p}) \in \dR \times \dR_+   \, ,
\end{equation}
and the corresponding multiplication law of the group $G$ reads
\begin{equation}\label{rob3}
(\tilde{q}^\prime, \tilde{p}^\prime)\cdot (\tilde{q}, \tilde{p})
= (\tilde{p}^\prime \tilde{q}
+ \tilde{q}^\prime, \tilde{p}^\prime \tilde{p} )\, .
\end{equation}
In the present paper we apply another simple parametrization, considered in
\cite{CA,AWT}, with the action of the group $G$ on $\dR_+$ defined as
\begin{equation}\label{rob4}
x'=(q,p)\cdot x = x/p + q,~~\mbox{ where }~~
(q,p) \in \dR \times \dR_+, .
\end{equation}
The corresponding multiplication law of the group is defined to be
\begin{equation}\label{rob3}
(q^\prime, p^\prime)\cdot (q, p) = (q/ p^\prime + q^\prime, p^\prime p )\, .
\end{equation}

The affine group $\Group{G}= \mathrm{Aff}(\dR)$ has two (nontrivial)
inequivalent irreducible unitary representations, \cite{Gel} and \cite{AK1,AK2},
defined in the Hilbert space $L^2 (\dR_+, d\nu(x))$, where
$d\nu(x):= dx/x$. In what follows, we choose the one defined by

\begin{equation}\label{m4}
U (q,p)\Psi (x) := e^{iqx} \Psi (x/p) \, ,
\end{equation}
where $\Psi \in L^2 (\dR_+, d\nu(x))$.

Integration over the affine group is defined as
\begin{equation}\label{m5}
 \int_G d\mu (q,p):=
\frac{1}{2\pi} \int_{-\infty}^\infty dq \int_0^\infty dp \, ,
\end{equation}
where the measure in \eqref{m5} is left invariant.

Any coherent state can be obtained as
\begin{equation}\label{m6}
 \langle x |q,p \rangle = U (q,p) \Phi (x),
\end{equation}
where $L^2 (\dR_+, d\nu (x))\ni \Phi (x) = \langle x | \Phi \rangle$, with
$\langle \Phi|\Phi \rangle=1$, is the so called fiducial vector.

The resolution of the identity in the Hilbert space $L^2 (\dR_+, d\nu(x))$
reads
\begin{equation}\label{m7}
 \int_G d\mu (q,p) | q,p\rangle  \langle q,p  | = A_\Phi \I \, ,
\end{equation}
where
\begin{equation}\label{m8}
 A_\Phi = \int_0^\infty \frac{dx}{x^2} |\Phi(x)|^2  < \infty \, .
\end{equation}
%
\subsection{Affine coherent states for the entire system}

Here, we again recall some essentials of the formalism of \cite{Gozdz:2018aai},
and insert suitable modifications resulting from the different parametrization
\eqref{rob4} of the affine group.

In the Cartesian product $\Pi=\Pi_1 \times \Pi_2$, the partial phase spaces
$\Pi_1$ or $\Pi_2$ are identified with the corresponding affine groups
$\Group{G}_1=\Group{Aff}_1(\dR)$ or $\Group{G}_2=\Group{Aff}_2(\dR)$. The
product of both affine groups $\Group{G}_{\Pi}= \Group{G}_1 \times \Group{G}_2 $
can be identified with the whole phase space and its action reads

\begin{equation}
\label{AffActPi}
\Pi \ni (\xi_1,\xi_2) \to \Ket{\xi_1,\xi_2 } = U(\xi_1,\xi_2)\Ket{\Phi}
:= U_1(\xi_1) \otimes U_2(\xi_2) \Ket{\Phi} \in \mathcal{H} \, ,
\end{equation}
where $\xi_k=(q_k,p_k)$ (with $k=1,2$), and where the entire Hilbert space is
the tensor product of two Hilbert spaces $\mathcal{H} = \mathcal{H}_1 \otimes
\mathcal{H}_2 = L^2(\dR_+ \times \dR_+,d\nu(x_1,x_2))$ with the measure
$d\nu(x_1,x_2)=d\nu(x_1)d\nu(x_2)$. The scalar product in $\mathcal{H}$ is
defined as
\begin{equation}
\label{ScalarProdL2Whole}
\BraKet{\psi_2}{\psi_1}= \int_0^\infty d\nu(x_1) \int_0^\infty d\nu(x_2)
\psi_1(x_1,x_2)^\star \psi_2(x_1,x_2) \, .
\end{equation}
The fiducial vector $\langle x_1, x_2|\Phi \rangle = \Phi(x_1,x_2)$ is a product
of two fiducial vectors $\Phi(x_1,x_2)= \Phi_1(x_1)\Phi_2(x_2)$. See
\cite{Gozdz:2018aai} for some subtleties concerning the choice of the vector
$\Phi$.

Finally, the explicit form of the action of the group $\Group{G}_{\Pi}$ on the
vector $\langle x_1, x_2 | \Psi \rangle = \Psi (x_1, x_2) \in \mathcal{H}$, in
the parametrization \eqref{rob4}, reads \cite{Gozdz:2018aai}:
\begin{equation}
\label{ActGPiH}
U(q_1,p_1,q_2,p_2)\Psi(x_1,x_2)= e^{iq_1x_1} e^{iq_2x_2}
\Psi(x_1/p_1,x_2/p_2)\, .
\end{equation}

\subsection{Quantum observables}

Making use of the resolution of identity in the Hilbert space $\mathcal{H}$, we
define the quantization of a classical observable $f$ defined in the phase space
$\Pi$ as follows \cite{Gozdz:2018aai}:
\begin{equation}\label{QuantObservPi}
 \hat{f}(t)= \frac{1}{A_{\Phi_1} A_{\Phi_2}}\int_{G_\Pi}  d\mu(\xi_1,\xi_2)
  \Ket{\xi_1,\xi_2}f(\xi_1,\xi_2)\Bra{\xi_1,\xi_2}\, ,
\end{equation}
where $ d\mu(\xi_1,\xi_2) := d\mu(q_1,p_1)d\mu(q_2,p_2)$.

We recommend \cite{Gozdz:2018aai} for a discussion of the properties of the
mapping that leads to \eqref{QuantObservPi}. If the operator $\hat{f}:
\mathcal{H} \rightarrow \mathcal{H}$ is unbounded, its possible self-adjoint
extensions require further examination \cite{Reed,Kon,GTV}.

\section{Quantum dynamics}

The mapping \eqref{QuantObservPi} applied to the classical Hamiltonian
reads
\begin{equation}\label{Qd1}
 \hat{H}(t)= \frac{1}{A_{\Phi_1} A_{\Phi_2}}\int_{G_\Pi}  d\mu(\xi_1,\xi_2)
  \Ket{\xi_1,\xi_2}H(t, \xi_1,\xi_2)\Bra{\xi_1,\xi_2}\, ,
\end{equation}
where $t$ is an evolution parameter of the classical level and where
\begin{equation}\label{Qd}
 \int_{G_\Pi}  d\mu(\xi_1,\xi_2) :=
\frac{1}{(2\pi)^2} \int_{-\infty}^{+\infty} dq_1\int_{-\infty}^{+\infty} dq_2
\int_0^{+\infty}dp_1\int_0^{+\infty}dp_2 \, .
\end{equation}

In our article \cite{Gozdz:2018aai} we apply the reduced phase space
quantization.  It means, we quantize the classical system with already resolved
dynamical constraint.  Its Hamilton's dynamics, corresponding to \eqref{rob1},
includes the generator of evolution in the physical phase space, i.e.  the
Hamiltonian $H$, and corresponding evolution parameter $t$.  As this Hamiltonian
system has no dynamical constraint, no quantum constraint occurs. This is quite
different from the Dirac quantization where the classical
constraint\footnote{For simplicity we assume there is only one constraint.} is
kept unsolved and is promoted to the quantum level so that it leads to an
operator type equation. The latter serves as the quantum transformation that
sometimes can be used to define a kind of quantum evolution, but in most cases
stays timeless (see \cite{MH} for more details).

As $H$ and $t$ is a classical canonical pair in \eqref{rob1}, it is reasonable
to assume that the quantum operator $\hat{H}$ corresponding to $H$ is a
generator of the evolution of the system in the Hilbert space
$\mathcal{H}$. More precisely, the operator $\hat{H}$ is the generator of
translations of the wave function of our quantum system with the corresponding
shift parameter $\tau$. It is natural to identify the classical shift
parameter $t$ and the quantum shift parameter $\tau$, i.e. we assume $\tau=t$.
This is a reasonable assumption as $t$ changes monotonically \cite{Gozdz:2018aai},
and it introduces the consistency between the classical and quantum levels.
Assuming the above identification of the evolution parameters, the translation
of the system from $t_0$ to $t$ is represented by the unitary operator $U(t,t_0)$
generated by $\hat{H}(t)$. The standard properties of the unitary evolution operators
in the Hilbert space $\mathcal{H}$:
\begin{eqnarray}
\label{eq:UnitaryEvOperator}
U(t,t)=1, \quad
U(t,t_0)^\dagger=U(t_0,t)=U(t,t_0)^{-1} \, , \nonumber \\
U(t_2,t_0)=U(t_2,t_1)U(t_1,t_0) \, ,
\end{eqnarray}
and continuity imply $\Psi(t)=U(t,t_0)\Psi(t_0)$.  It further means that the
quantum evolution of our gravitational system can be equivalently defined by the
Schr\"{o}dinger type equation:
\begin{equation}\label{Qd2}
  i  \frac{\partial}{\partial t}|\Psi (t) \rangle
= \hat{H}(t) |\Psi (t) \rangle \; .
\end{equation}
The classical time $t$ occurs in \eqref{Qd2} because it enters the integrand of
\eqref{Qd1}. We do not quantize the classical time $t$.  In the case $t$ were a
quantum observable, it would be mapped into a quantum operator \cite{AGMGAP},
but we do not consider here such a case.


\subsection{Classical dynamics near the singularity}

Near the gravitational singularity, the terms $\exp(2q_1)$ and $\exp(q_2 - q_1)$
in the function $F$ can be neglected (see \cite{Gozdz:2018aai} for more details)
so that we have
\begin{equation}\label{ns1}
F(t, q_1,q_2,p_1,p_2) \longrightarrow F_0 (t, p_1,p_2) :=
p_1 p_2-\frac{1}{4} (t-p_1-p_2)^2 \, .
\end{equation}
This form of $F$ leads to the simplified form of the Hamiltonian \eqref{vis1a}
which now reads
\begin{eqnarray} \label{ns2a}
&& H_0 (t,q_2,p_1, p_2) :=
\begin{cases}
- q_2 - \ln F_0(t,p_1, p_2),~~~
\text{for}~~~F_0(t,p_1, p_2) >0 \\
0,~~~\text{for}~~F_0(t,p_1, p_2) < 0
\end{cases}
\end{eqnarray}
with $\lim_{F_0\rightarrow 0^-} H_0 = 0~~\text{and}~~\lim_{F_0\rightarrow 0^+}
H_0 = + \infty $.  In fact, the condition
\begin{equation}\label{regF3a}
  F_0 (t, p_1,p_2) > 0
\end{equation}
defines the available part of the physical phase space $\Pi$ for the classical
dynamics, defined by \eqref{kps2}, which corresponds to the approximation
\eqref{ns1}.  Eqs. \!\eqref{ns1}--\eqref{ns2a} define the approximation to our
original Hamiltonian system describing the dynamics in the close vicinity of the
singularity.

\subsection{Quantum dynamics near the singularity}

Calculations similar to the ones carried out in our paper \cite{Gozdz:2018aai},
applied to the Hamiltonian \eqref{ns2a}, lead to the Schr\"{o}dinger equation
\eqref{Qd2} in the form:
\begin{equation}\label{ns9}
i \frac{\partial}{\partial t} \Psi(t,x_1,x_2) =
 \left( i \frac{\partial }{\partial x_2}  -\frac{i}{2x_2} - \tilde{K}(t,x_1,x_2)
 \right) \Psi(t,x_1,x_2)\, ,
\end{equation}
where $\Psi(t,x_1,x_2):= \langle x_1, x_2 | \Psi (t) \rangle$. The function
$\tilde{K}$ reads
\begin{equation} \label{ns6}
\tilde{K} (t, x_1, x_2) :=
\frac{1}{A_{\Phi_1} A_{\Phi_2}}\; \int_0^\infty d p_1 \int_0^\infty d p_2\,
\widetilde{\ln}\big(F_0(t, p_1,p_2)\big)
|\Phi_1(x_1/p_1 )|^2 |\Phi_2(x_2/p_2 )|^2 \, ,
\end{equation}
where
\begin{eqnarray} \label{tildeln}
&& \widetilde{\ln}\big(F_0(t, p_1, p_2)\big) :=
\begin{cases}
\ln\big(F_0(t, p_1, p_2 \big),~~~
\text{for}~~~ F_0(t,p_1, p_2 ) >0 \\
0,~~~\text{for}~~~F_0(t,p_1, p_2)  < 0
\end{cases}
\end{eqnarray}
The fiducial function $\Phi_2 (x) \in \dR$ should satisfy the conditions:
\begin{equation}\label{ns7}
\Phi_2 (x) =: x \tilde{\Phi}(x),~~~ \lim_{x \rightarrow 0^+}\tilde{\Phi}(x) = 0,~~~\lim_{x \rightarrow +\infty}\tilde{\Phi}(x) = 0\, ,
\end{equation}
and the solution to \eqref{ns9} is expected to have the properties:
\begin{equation}\label{ns8}
\Psi (t, x_1,x_2) =:
\sqrt{x_2} \,\tilde{\Psi}(t, x_1,x_2),~~~\lim_{x_2
\rightarrow 0^+}\tilde{\Psi}(t, x_1,x_2) = 0,
~~~\lim_{x_2 \rightarrow +\infty}\tilde{\Psi}(t, x_1,x_2) = 0 \, .
\end{equation}

The mathematical structure of Eq. \!\eqref{ns9} is similar to the corresponding
one of the paper \cite{Gozdz:2018aai}.  The difference concerns just one part of
these equations, namely the actual function $\tilde{K} (t, x_1, x_2)$  and
the function $K (t, x_1, x_2)$ in \cite{Gozdz:2018aai}. Therefore, the general
solution to \eqref{ns9} reads
\begin{equation}\label{ns10}
\Psi(t,x_1,x_2)=\eta(x_1,x_2+t-t_0)\, \sqrt{\frac{x_2}{x_2+t-t_0}}
\, \exp\left(i \int_{t_0}^t \tilde{K}(t',x_1,x_2+t-t')\,dt' \right) \, ,
\end{equation}
where $ t \ge t_0 >0$, and where $\eta (x_1, x_2):= \Psi (t_0, x_1,x_2)$ is the
initial state satisfying the condition
\begin{equation}\label{nor2}
 \eta(x_1,x_2) = 0~~~~\mbox{for}~~~~x_2 < t_H \, ,
\end{equation}
with $t_H > 0$ being the parameter of our model. This condition is consistent
with \eqref{ns8} and for $t < t_H$ we get (see \cite{Gozdz:2018aai})
\begin{equation} \label{nor3}
\BraKet{\Psi(t)}{\Psi(t)} =
\int_0^\infty \frac{dx_1}{x_1}  \int_{t_H}^\infty \frac{d{x_2}}{x_2}\,
|\eta(x_1,x_2)|^2 \,,
\end{equation}
so that the inner product is time independent, which implies that the quantum
evolution is unitary. Due to \eqref{nor2}, the probability of finding the system
in the region with $x_2 < t_H$ vanishes so that this region does not contribute
to the expectation values of observables. These results are consistent with the
results of \cite{Gozdz:2018aai}.

Since the mathematical structure of the dynamics presented here and in
\cite{Gozdz:2018aai} are quite similar, the operation of time reversal turns
\eqref{ns9} into the equation:
\begin{equation}\label{qb2}
i \frac{\partial}{\partial t} \tilde{\Psi}(t,x_1,x_2) =
 \left(- i \frac{\partial }{\partial x_2}  +\frac{i}{2x_2}
- \tilde{K}(-t,t,x_1,x_2) \right) \tilde{\Psi}(t,x_1,x_2)\, ,
\end{equation}
where $\tilde{\Psi} (t, x_1, x_2) := \Psi (- t, x_1, x_2)^\ast$. Consequently,
the solution to \eqref{qb2} for $t < 0$, reads
\begin{equation}\label{qb3}
\tilde{\Psi}(t,x_1,x_2)=
\eta(x_1,x_2 + |t| - |t_0|)\, \sqrt{\frac{x_2}{x_2 + |t| - |t_0|}}\,
\exp\left(i \int_{t_0}^t \tilde{K}(-t',x_1,x_2-t+t')\,dt' \right) \, ,
\end{equation}
where $ |t|\geq |t_0|$, and where $\eta (x_1, x_2):= \tilde{\Psi} (t_0,
x_1,x_2)$ is the initial state.

The unitarity of the evolution (with $t_0 = 0$) can be obtained again if
\begin{equation}\label{qb4}
 \eta(x_1,x_2) = 0~~~~\mbox{for}~~~~x_2 < |t_H| \, ,
\end{equation}
which corresponds to the condition \eqref{nor2}.

Since the solutions \eqref{ns10} and \eqref{qb3} differ only by the
corresponding phases, the probability density is continuous at $t=0$, which
means that we are dealing with quantum bounce at $t=0$ (that marks the classical
singularity).

\section{Conclusions}

The quantum dynamics we have obtained does not depend essentially on the applied
parametrization of the affine group. Two different parametrizations give
qualitatively the same results, which differ only slightly quantitatively. The
latter is meaningless if we only insist on the main result which is the
resolution of the classical singularity.

We have applied the simplest two group parametrizations. The general one can be
presented in the form of the action of the group on $\dR_+$ as follows
\cite{AWT}:
\begin{equation}\label{genr}
  \dR_+ \ni x \rightarrow x^\prime = \xi (p,q)\cdot x
+ \eta (p,q) \cdot p \in \dR_+ \, .
\end{equation}
We expect that in the case $ \xi (p,q) = \tilde{\xi}(q)$ and $\eta (p,q) =
\tilde{\eta}(p)$, the result of quantization will be qualitatively the same as
the one obtained in the present paper.

The effect of using quite general parametrization considered in \cite{AWT},
applied to the quantization of our gravitational system, would need separate
examination and is beyond the scope of the present paper.  We may stay with the
simplest parametrizations if we do not test the quantization method as such, but
intend to get the result with satisfactory physics. After experimental or
observational data on quantum gravity become available, the way of choosing the
most suitable group parametrization will obtain sound guideline.



\begin{thebibliography}{99}


\bibitem{Gozdz:2018aai}
  A.~ G\'{o}\'{z}d\'{z}, W.~Piechocki and G.~Plewa,
  ``Quantum Belinski-Khalatnikov-Lifshitz scenario,''
  Eur.\ Phys.\ J.\ C {\bf 79},  45 (2019).


\bibitem{AWT}
 A.~ G\'{o}\'{z}d\'{z}, W.~Piechocki and T.~Schmitz,
``On the dependence of the affine coherent states quantization on the parametrization of the affine group'',
arXiv: 1908.100039 [math-ph].


\bibitem{BKL22} V. A. Belinskii, I. M. Khalatnikov, and E. M. Lifshitz,
  ``Oscillatory approach to a singular point in the relativistic cosmology'',
  Adv. Phys. {\bf 19}, 525 (1970).

\bibitem{BKL33} V. A. Belinskii, I. M. Khalatnikov, and E. M. Lifshitz, ``A
  general solution of the Einstein equations with a time singularity'',
  Adv. Phys.  {\bf 31}, 639 (1982).

\bibitem{Bel} V.~Belinski and M.~Henneaux, {\em The Cosmological
    Singularity} (Cambridge University Press, Cambridge, 2017).

 \bibitem{ryan} V. A. Belinskii, I. M. Khalatnikov, and M. P. Ryan,
``The oscillatory regime near the singularity in Bianchi-type IX
universes'',  Preprint {\bf 469} (1971), Landau Institute for
Theoretical Physics, Moscow (unpublished); published as Secs. 1
and 2 in  M. P. Ryan, Ann. Phys. {\bf 70},  301 (1971).

 \bibitem{Nick}
  C.~Kiefer, N.~Kwidzinski, and W.~Piechocki,
  ``On the dynamics of the general Bianchi IX spacetime near the singularity'',
 Eur. Phys. J. C {\bf 78}, 691 (2018).

\bibitem{Czuchry:2012ad}
  E.~Czuchry and W.~Piechocki,
  ``Bianchi IX model: Reducing phase space,''
  Phys.\ Rev.\ D {\bf 87},  084021 (2013).

\bibitem{CA} C. R. Almeida, H. Bergeron, J.-P. Gazeau, and A. C. Scardua,
``Three examples of quantum dynamics on the half-line with smooth bouncing'',
Annals of Physics {\bf 392}, 206 (2018).

\bibitem{Gel} I. M. Gel$'$fand and M. A. Na\"{i}mark, ``Unitary representations of the group of linear transformations of the
straight line'', Dokl. Akad. Nauk. SSSR {\bf 55}, 567 (1947).

\bibitem{AK1} E. W. Aslaksen and J. R. Klauder, ``Unitary
Representations of the Affine Group'', J. Math. Phys. {\bf 9}, 206
(1968).

\bibitem{AK2} E. W. Aslaksen and J. R. Klauder, ``Continuous Representation
Theory Using Unitary Affine Group'', J. Math. Phys. {\bf 10}, 2267 (1969).

\bibitem{Reed} M. Reed and B. Simon, {\it Methods of Modern Mathematical
    Physics} (San Diego, Academic Press, 1980), Vols I and II.

\bibitem{Kon} K. Schm\"{u}dgen, {\it Unbounded Self-adjoint Operatotrs on Hilbert Space}
(San Francisko, Springer, 2012).

\bibitem{GTV} D. M. Gitman, I. V. Tyutin, and B. L. Voronov, {\it Self-adjoint Extensions
in Quantum Mechanics} (Springer, Birkh\"{a}user, 2012).

\bibitem{MH} M. Henneaux and C. Teitelboim, {\it Quantization of Gauge Systems } (Princeton
University Press, Princeton, 1992).

\bibitem{AGMGAP}   A.~ G\'{o}\'{z}d\'{z},  M.~ G\'{o}\'{z}d\'{z}, and A.~P\c{e}drak, ``Projection
evolution of quantum states. Part I.'', arXiv:1910.111987 [quant-ph].

\end{thebibliography}
\end{document}